\documentclass[11pt,a4paper]{article}

\usepackage[utf8]{inputenc}
\usepackage[T1]{fontenc}
\usepackage{amsmath,amssymb}
\usepackage{geometry}
\usepackage{booktabs}
\usepackage{enumitem}
\usepackage{hyperref}
\usepackage{cleveref}
\usepackage{graphicx}
\usepackage{xcolor}
\usepackage{titlesec}
\usepackage[numbers,sort&compress]{natbib}

\hypersetup{colorlinks=true,linkcolor=blue!60!black,citecolor=green!50!black,urlcolor=blue!70!black}

\title{CASCADE: Cross-scale Advective Super-resolution\\
       with Climate Assimilation and Downscaling Evolution\\[0.5em]
       }
\author{Alexander Kovalenko}
\date{}

\begin{document}
\maketitle

\begin{abstract}
Super-resolution of geophysical fields presents unique challenges beyond natural image enhancement: fine-scale structures must respect physical dynamics, conserve mass and energy, and evolve coherently in time. These constraints are especially critical for extreme events, where rare, localized, high-intensity features drive impacts and where temporally inconsistent ``hallucinated'' detail can misrepresent hazards. We introduce CASCADE (Cross-scale Advective Super-resolution with Climate Assimilation and Downscaling Evolution), a framework that reframes spatiotemporal super-resolution as an explicit transport process across scales. Rather than hallucinating high-frequency content per pixel, CASCADE reconstructs fine structure by iteratively advecting coarse information along learned, flow-conditioned velocity fields through semi-Lagrangian warping. The architecture decomposes motion into resolved (large-scale) and subgrid (unresolved) components, mirroring the closure problem in numerical weather prediction, and enforces low-resolution consistency through an assimilation-style innovation step. Evaluated on SEVIR radar data for 4$\times$ super-resolution of severe convective storms, CASCADE outperforms strong baselines across both continuous metrics (PSNR, SSIM, MAE) and threshold-based skill scores (CSI, HSS, POD) while providing interpretable diagnostics through visualizable velocity and correction fields. By encoding advection as the fundamental operator rather than learning it implicitly, CASCADE produces temporally coherent, physically consistent, and mass-conserving reconstructions well suited to advection-dominated extremes in atmospheric and oceanic applications.
\end{abstract}

\section{Introduction}

Over the past decade, deep neural networks have achieved remarkable 
performance across perception and language tasks by learning multiscale 
representations from data. These advances have naturally extended to 
scientific computing, where machine learning (ML) models increasingly serve 
as fast, accurate surrogates for traditional numerical methods and as 
components in hybrid modeling pipelines.

In atmospheric science, ML-based models are now used both as rapid 
approximations to numerical weather prediction (NWP) systems and as 
post-processing tools. Recent global deep-learning forecasters, including 
FourCastNet \citep{pathak2022fourcastnet}, Pangu-Weather 
\citep{bi2023panguweather}, and GraphCast \citep{lam2023graphcast}, demonstrate 
that training on large reanalysis archives can produce medium-range forecasts 
with skill competitive to traditional NWP at a fraction of the computational 
cost. Complementary efforts explore foundation-model approaches that unify 
tasks across weather and climate domains. For instance, ClimaX pretrains a 
transformer on climate-model output and successfully transfers learned 
representations across multiple variables, lead times, and downstream tasks 
\citep{nguyen2023climax}.

Despite these advances, extreme weather events remain a major challenge. 
Extremes are rare, strongly nonlinear, and often controlled by intermittent 
small-scale processes that are difficult to represent in coarse-resolution 
datasets. Standard training objectives such as mean-squared error tend to 
bias models toward smooth conditional means, systematically under-representing 
the tails of the distribution that matter most for hazard assessment and 
emergency response. This has motivated growing interest in models specifically 
designed to capture extremes, including high-resolution precipitation forecasting 
systems and nowcasting architectures such as MetNet-3 \citep{merchant2023metnet3}.

To improve both robustness and physical consistency, physics-informed neural 
networks (PINNs) and related physics-guided ML approaches incorporate known 
structure, differential constraints, conservation laws, symmetries, directly 
into the learning process \citep{karniadakis2021pinns,cuomo2022scientificml}. 
While scaling classical PINN formulations to high-dimensional chaotic flows 
remains challenging, the underlying principle is sound: \emph{encode the right 
operators} rather than forcing networks to rediscover fundamental physics from 
scratch.

Super-resolution (SR) methods developed for natural images typically prioritize 
perceptual realism, accepting that multiple high-frequency completions may be 
visually plausible for a given low-resolution input. Geophysical SR, by contrast, 
faces stricter constraints. Fine-scale structures must be consistent with both 
the coarse-scale state \emph{and} the subsequent spatiotemporal evolution. A 
visually convincing precipitation filament is scientifically meaningless if it 
advects in the wrong direction, violates mass conservation, or fails to match 
coarse-scale observations when averaged. These requirements motivate SR approaches 
that explicitly incorporate temporal context, physical transport mechanisms, and 
assimilation-style constraints rather than relying solely on per-pixel 
hallucination.

For extreme events, these requirements are non-negotiable: impact-relevant
quantities (e.g., peak precipitation rates, storm cores, sharp fronts) are
precisely the high-frequency structures most likely to be smoothed by regression
or spuriously created by purely generative detail. A useful downscaling method
must therefore (i) move intense features in physically plausible directions,
(ii) preserve their integrated magnitude as they advect, and (iii) remain
consistent with the observed coarse-scale state. CASCADE is designed around
these constraints by making transport the core operator and using an
assimilation-style correction to prevent drift.

For extreme events, these requirements are non-negotiable: impact-relevant
quantities (e.g., peak precipitation rates, storm cores, sharp fronts) are
precisely the high-frequency structures most likely to be smoothed by regression
or spuriously created by purely generative detail. A useful downscaling method
must therefore (i) move intense features in physically plausible directions,
(ii) preserve their integrated magnitude as they advect, and (iii) remain
consistent with the observed coarse-scale state. CASCADE is designed around
these constraints by making transport the core operator and using an
assimilation-style correction to prevent drift.

This paper introduces CASCADE (\textbf{C}ross-scale \textbf{A}dvective 
\textbf{S}uper-resolution with \textbf{C}limate \textbf{A}ssimilation and 
\textbf{D}ownscaling \textbf{E}volution), a framework for super-resolution 
and dynamical downscaling of spatiotemporal geophysical fields. CASCADE 
reconstructs fine-scale structure through iterative, flow-guided semi-Lagrangian 
advection across scales. In its dynamical downscaling variant, it incorporates 
an explicit innovation step analogous to data assimilation, ensuring temporal 
consistency while respecting coarse-scale observations.

\section{Related Work}

\subsection{Super-Resolution in Computer Vision}

Super-resolution has a rich history in computer vision, where the goal is to 
reconstruct high-resolution images from low-resolution observations. Classical 
deep SR methods achieved rapid progress through CNN-based residual architectures 
such as EDSR \citep{lim2017edsr} and attention-based designs such as RCAN 
\citep{zhang2018rcan}. For perceptual quality, adversarial and perceptual-loss 
formulations, exemplified by ESRGAN \citep{wang2018esrgan}, can generate 
visually sharp textures. However, the high-frequency content produced by these 
methods is not uniquely determined by the input; multiple plausible completions 
exist.

This ambiguity poses problems for geophysical applications. Downscaled atmospheric 
or oceanic fields are evaluated not only by pixel-wise error but also by their 
spectral characteristics, representation of extremes, and spatiotemporal consistency. 
A sharp gradient that looks convincing in isolation may be physically meaningless 
if it appears at the wrong location, moves in the wrong direction, or violates 
conservation principles. These requirements motivate methods that incorporate 
dynamics and physical constraints rather than purely appearance-driven texture 
synthesis.

Recent generative approaches (e.g., diffusion-based nowcasting and downscaling)
have improved perceptual fidelity and can represent forecast uncertainty, but
they still require strong constraints to ensure transport-consistent evolution
and physically meaningful extremes \citep{leinonen2023ldmnowcasting,ling2024srndiff,wen2024duocast,foo2025stldm}.

\subsection{Deep Learning for Climate and Weather Downscaling}

In climate and weather applications, deep learning has been widely adopted as 
a form of statistical downscaling. Early work explicitly adapted image SR 
architectures to climate data, with DeepSD \citep{vandal2017deepsd} among the 
pioneering examples. Since then, a substantial body of research has explored 
CNNs, U-Nets, GANs, and transformers for meteorological SR across different 
variables, regions, and resolutions. Systematic reviews have identified recurring 
challenges including scale-factor dependence, multivariate consistency, uncertainty 
quantification, and out-of-distribution generalization 
\citep{huang2023systematicreview}.

Recent probabilistic approaches reframe downscaling as distribution matching, 
using generative models to explicitly target tail behavior and multivariate 
dependencies \citep{wan2024genbcsr}. These advances are complementary to CASCADE. 
While probabilistic models focus on capturing the full conditional distribution 
of fine-scale fields, CASCADE emphasizes making reconstructions 
\emph{transport-consistent} across time. Rather than regenerating fine-scale 
structure independently at each time step, CASCADE advects it forward coherently, 
ensuring that recovered structures evolve in accordance with observed motion.

\subsection{Neural Operators and Physics-Informed Learning}

Neural operators offer a distinct perspective by learning mappings between 
function spaces, aiming to approximate solution operators of partial differential 
equations (PDEs) in a resolution-independent manner 
\citep{azizzadenesheli2024neuraloperators}. Fourier Neural Operators (FNOs) 
\citep{li2020fno} and physics-informed neural operators (PINOs) \citep{li2021pino} 
have demonstrated strong performance as fast surrogates for parametric PDE 
families. They have been successfully deployed as building blocks in large-scale 
forecasting systems such as FourCastNet \citep{pathak2022fourcastnet}.

The advantages of neural operators include global receptive fields, flexible 
resolution transfer, and differentiability for inverse problems. Challenges 
include substantial data requirements, potential brittleness under domain shift, 
and architectural constraints tied to geometry and discretization (though recent 
variants address irregular domains and multiresolution settings). CASCADE is 
compatible with this operator-learning perspective but targets a different 
design point. Rather than learning the full spatiotemporal evolution operator 
end-to-end, CASCADE isolates \emph{advection}, the dominant transport 
mechanism, as an explicit, differentiable operation. The network learns only 
the unresolved subgrid correction that drives cross-scale refinement, reducing 
what must be learned from data and improving physical interpretability.

Physics-informed neural networks (PINNs) and related physics-guided approaches 
provide yet another avenue for incorporating domain knowledge 
\citep{karniadakis2021pinns,cuomo2022scientificml}. By embedding PDE residuals, 
conservation laws, and symmetries directly into loss functions or architectures, 
PINNs are particularly attractive when observations are sparse and when physical 
constraints must hold beyond the training distribution. However, they can be 
challenging to optimize and difficult to scale to turbulent, multiscale 
spatiotemporal systems.

CASCADE follows a similar philosophy, encode the right physics, but does so 
primarily through architectural choices rather than loss-function regularization. 
Semi-Lagrangian advection is built into the model as the fundamental mechanism 
for transporting information in space and time. Assimilation-style innovation 
enforces consistency with coarse observations. This structural approach makes 
the physics explicit and interpretable while maintaining end-to-end 
differentiability for efficient training.

\subsection{Positioning CASCADE}

Taken together, prior work reveals a three-way trade-off between expressiveness 
(purely data-driven SR and generative models), physical fidelity (PINNs and 
hard physics constraints), and scalability (operator learning and fast global 
architectures). CASCADE is designed to navigate this space by:

\begin{itemize}
  \item Replacing per-pixel hallucination with explicit transport across scales, 
        ensuring that fine-scale structure is moved rather than invented;
  \item Decomposing dynamics into resolved (large-scale) and subgrid (unresolved) 
        components, mirroring the closure problem in numerical modeling;
  \item Coupling SR with temporal evolution and assimilation-style correction, 
        producing reconstructions that are temporally coherent and observationally 
        consistent.
\end{itemize}

The resulting framework is physically interpretable, temporally coherent, and 
naturally suited to advection-dominated geophysical fields where temporal 
sequences and paired coarse-fine training data are available.

\section{Methodology}
\label{sec:methodology}

\subsection{Overview}

We introduce CASCADE (\textbf{C}ross-scale \textbf{A}dvective
\textbf{S}uper-resolution with \textbf{C}limate \textbf{A}ssimilation and
\textbf{D}ownscaling \textbf{E}volution), a framework for
super-resolution (SR) of spatiotemporal geophysical fields rooted in
a single physics-informed principle: \emph{detail is not hallucinated
pixel by pixel but reconstructed by learned transport across scales}.
The name reflects the physical analogy: just as turbulent energy cascades
from large to small scales, CASCADE reconstructs fine-scale structure from
coarse fields through iterative, flow-guided transport steps, an
\emph{inverse cascade} in scale space.

The main contribution is to reframe geophysical SR as an \emph{explicit
transport process across scales}.  Rather than predicting high-frequency
content per pixel, CASCADE reconstructs fine structure by iteratively
advecting information with learned, flow-conditioned velocity fields,
while enforcing low-resolution consistency and (in the dynamical variant)
an assimilation-style innovation update.  By hard-wiring transport into
the core operator, the method produces reconstructions that are (i)
interpretable via visualizable velocity and correction fields, and (ii)
naturally temporally coherent.

We consider two architecture variants:

\begin{description}[style=nextline,leftmargin=2em]
  \item[CASCADE-SR (Problem~A --- spatial SR with temporal context)]
    Given $N$ consecutive low-resolution (LR) frames, produce a single
    high-resolution (HR) snapshot of the last frame.
    A \textsc{FlowNet} estimates resolved motion from the full temporal
    window; a \textsc{SubgridNet} iteratively warps the bilinear-upsampled
    frame along a learned subgrid velocity conditioned on the resolved flow
    and local gradients:
    \begin{equation}\label{eq:scale-step}
      u_{s+1}(\mathbf{x})
        = u_s\!\bigl(\mathbf{x} - \mathbf{v}_s(\mathbf{x})\bigr),
      \qquad
      \mathbf{v}_s
        = \textsc{SubgridNet}\!\bigl(u_s,\;\mathbf{v}_t,\;\nabla u_s,\;s/S\bigr).
    \end{equation}

  \item[CASCADE-DD (Problem~B --- full dynamical downscaling)]
    Given $N$ consecutive LR frames, produce the \emph{entire} HR
    sequence by time-stepping through the observations.  At every time
    step $t$ the HR state is (i)~advected forward by a learned inter-frame
    flow, (ii)~corrected by an \textsc{AssimilationNet} that compares the
    forecast with the LR observation (innovation), and (iii)~refined by
    the \textsc{SubgridNet} scale-evolution loop:
    \begin{align}
      \mathbf{v}_t        &= \textsc{FlowNet}(u_{t-1},\;\widetilde{\mathrm{LR}}_t),
                          \label{eq:dd-flow}\\
      u_{\mathrm{adv}}    &= \mathrm{warp}(u_{t-1},\;\mathbf{v}_t),
                          \label{eq:dd-adv}\\
      \delta              &= \textsc{AssimNet}\!\bigl(
                              u_{\mathrm{adv}},\;
                              \widetilde{\mathrm{LR}}_t,\;
                              \widetilde{\mathrm{LR}}_t - \widetilde{\mathrm{pool}(u_{\mathrm{adv}})},\;
                              \mathbf{v}_t\bigr),
                          \label{eq:dd-assim}\\
      u_t                 &= \textsc{ScaleRefine}\!\bigl(
                              u_{\mathrm{adv}} + \delta,\;\mathbf{v}_t\bigr).
                          \label{eq:dd-refine}
    \end{align}
    Here $\widetilde{(\cdot)}$ denotes bilinear lifting to the HR grid and
    $\mathrm{pool}(\cdot)$ is average-pooling back to LR.
\end{description}

Both architectures use semi-Lagrangian advection (differentiable
\texttt{grid\_sample}) as the core operation instead of per-pixel
regression, and both are trained end-to-end with pixel loss,
LR-consistency loss, and velocity-smoothness regularisation.

\subsection{The advection equation: why transport is the right inductive bias}
\label{sec:advection}

The evolution of nearly every atmospheric and oceanic scalar field
$u(\mathbf{x},t)$---temperature, moisture, tracer concentration, vorticity---is
governed at leading order by the \emph{advection equation}:
\begin{equation}\label{eq:advection}
  \frac{\partial u}{\partial t}
  + \mathbf{v}\cdot\nabla u
  = \mathcal{S},
\end{equation}
where $\mathbf{v}(\mathbf{x},t)$ is the wind (or current) field and
$\mathcal{S}$ collects sources, sinks, and diffusion.  In the absence
of sources ($\mathcal{S}=0$), the field is simply carried along by the
flow: no new values are created or destroyed, existing values are
\emph{moved}.

This is the central insight behind CASCADE.  When a coarse-resolution
field is missing fine-scale detail, much of that detail was not
destroyed, it was \emph{averaged away} during the coarsening
(filtering) process.  A temperature front that is 10\,km wide in
reality becomes a 50\,km gradient smear at ERA5 resolution.  The
information about the front's existence is still present in the
coarse field (as a weak gradient); what is missing is the spatial
\emph{concentration} of that gradient.

CASCADE recovers sharp structure by learning a velocity field that
re-concentrates the coarse gradients, the same operation that the
real atmosphere performs continuously via advective deformation.
The semi-Lagrangian warp
\begin{equation}\label{eq:warp}
  u_{s+1}(\mathbf{x}) = u_s\bigl(\mathbf{x} - \mathbf{v}_s(\mathbf{x})\bigr)
\end{equation}
is a discrete approximation to solving \cref{eq:advection} backward
along characteristics.  It is unconditionally stable (no CFL
condition), differentiable (via \texttt{grid\_sample}), and
conservative up to interpolation error.

\paragraph{Why not just use a CNN decoder?}
A convolutional decoder must learn, from pixel intensities alone,
both \emph{what} features to sharpen and \emph{where} to sharpen
them.  By contrast, the warp operation factorises the problem:
the network only needs to predict the velocity field $\mathbf{v}$
(a 2D vector per pixel), and the sharpening follows automatically
from the transport.  This reduces the effective dimensionality
of what the network must learn and biases it toward physically
plausible solutions.

\subsection{Frontogenesis: how advection creates fine-scale structure}
\label{sec:frontogenesis}

The physical process that most directly motivates CASCADE is
\emph{frontogenesis}, the creation and intensification of sharp
gradients (fronts) by large-scale deformation.

Consider a smooth temperature field $\theta(\mathbf{x})$ with a
weak gradient $|\nabla\theta|$.  If the wind has a component of
convergence along the gradient direction, the gradient tightens
over time.  The frontogenesis function is:
\begin{equation}\label{eq:frontogenesis}
  F = \frac{D}{Dt}|\nabla\theta|
  = -|\nabla\theta|\,
    \bigl(E\cos 2\beta - D_{\mathrm{div}}\bigr),
\end{equation}
where $E$ is the total deformation rate, $\beta$ is the angle
between the axis of dilatation and the gradient vector, and
$D_{\mathrm{div}}$ is the divergence.  When the deformation
is oriented to compress the isotherms ($F > 0$), the gradient
sharpens exponentially.

This is \emph{exactly} what the SubgridNet learns to do.
Given a bilinear-upsampled field (weak, smooth gradients), the
SubgridNet predicts a velocity $\mathbf{v}_s$ whose convergence
is aligned with existing gradients, compressing them into sharper
features.  The network does not need to ``imagine'' where fronts
should be, it reads the gradient direction from $\nabla u_s$
(provided explicitly as input) and learns the deformation pattern
that concentrates it.

This is why the SubgridNet receives the local gradients
$\partial u/\partial x$ and $\partial u/\partial y$ as input
channels: they tell the network \emph{where} gradients exist and
in which direction, so it can predict the convergent velocity that
sharpens them.  A standard CNN decoder has no such explicit gradient
signal and must discover this structure implicitly.

\subsection{Scale decomposition and subgrid closure}
\label{sec:subgrid}

In numerical weather prediction and climate modelling, the governing
equations are solved on a finite grid with spacing $\Delta x$.  Processes
smaller than $\sim\!2\Delta x$ are unresolved and must be
\emph{parameterised}, represented as functions of the resolved-scale
variables.  This is the \emph{closure problem}: the evolution of the
resolved scales depends on the unresolved scales, so one must model
the effect of the small on the large.

CASCADE mirrors this decomposition directly:
\begin{itemize}
  \item The \textbf{resolved flow} $\mathbf{v}_t$ (from FlowNet)
    corresponds to the large-scale, grid-resolved advection, the
    wind field that a coarse model can represent.  In ERA5 at
    0.25$^\circ$, this is synoptic-scale motion: cyclone
    propagation, jet stream steering, frontal advection.

  \item The \textbf{subgrid velocity} $\mathbf{v}_s$ (from SubgridNet)
    corresponds to the parameterised transport by unresolved
    processes: convective updrafts, orographic flow splitting,
    sea-breeze convergence, turbulent mixing, all the motions
    that exist below the LR grid scale but shape the fine-scale
    structure of the field.
\end{itemize}

The separation is not prescribed by hand; it is \emph{learned} from
paired LR--HR data.  However, the architecture \emph{enforces} that
the decomposition exists: FlowNet and SubgridNet are separate networks
with different inputs, capacities, and velocity scales.  This
structural constraint is a form of inductive bias: it tells the model
``there are two kinds of transport operating at different scales,''
which is true for virtually all geophysical fields.

\paragraph{Connection to Reynolds averaging.}
If we denote the full (HR) velocity as
$\mathbf{u} = \bar{\mathbf{u}} + \mathbf{u}'$, where $\bar{\mathbf{u}}$
is the spatially filtered (LR-resolvable) part and $\mathbf{u}'$ is the
fluctuation, then the filtered advection equation contains the
\emph{Reynolds stress} term
$\nabla\cdot(\overline{\mathbf{u}'\theta'})$ that requires closure.
Classical parameterisations model this as diffusion (eddy viscosity)
or mass-flux schemes.  CASCADE's SubgridNet is a learned,
spatially varying, nonlinear closure: it predicts the \emph{effective
velocity} of the subgrid transport conditioned on the local resolved
state, the resolved flow, and the gradient structure.

\subsection{Conservation: why warp preserves what regression does not}
\label{sec:conservation}

The continuity equation for a conserved tracer in an incompressible
flow is:
\begin{equation}\label{eq:continuity}
  \frac{\partial u}{\partial t}
  + \nabla\cdot(u\,\mathbf{v}) = 0
  \qquad\Longrightarrow\qquad
  \int u\,dA = \mathrm{const.}
\end{equation}
The integral of $u$ over the domain does not change, mass (or energy,
or moisture) is conserved.

A semi-Lagrangian warp (\cref{eq:warp}) approximates this: it remaps
existing field values to new locations without creating or destroying
intensity.  The total integral changes only due to boundary effects
(values that leave or enter the domain) and interpolation error, both
of which are small.  By contrast, a CNN decoder generates each output
pixel independently; there is no architectural reason for the sum of
outputs to match the sum of inputs.

CASCADE further enforces conservation through the \textbf{LR-consistency
loss}:
\begin{equation}\label{eq:lr-loss}
  \mathcal{L}_{\mathrm{LR}}
  = \bigl\|\mathrm{pool}(\hat{u}_{\mathrm{HR}}) - u_{\mathrm{LR}}\bigr\|^2.
\end{equation}
This guarantees that the HR output, when averaged back to the LR grid,
recovers the original coarse observation.  Physically, this means the
super-resolved field is a \emph{valid disaggregation} of the coarse
field: it adds spatial detail but does not change the area-average.
For precipitation this ensures that the total rainfall over a grid box
is preserved; for temperature it ensures the area-mean is correct.

\subsection{Data assimilation: why the time-stepping loop is a
            learned analysis cycle}
\label{sec:assimilation}

In operational weather forecasting, the analysis cycle alternates
between two steps:
\begin{enumerate}
  \item \textbf{Forecast:} Run the model forward from the previous
    analysis $u_{t-1}^a$ to produce a background (prior) $u_t^b$.
  \item \textbf{Analysis:} Correct the background toward the new
    observations $y_t$ to produce the analysis:
    \begin{equation}\label{eq:kalman}
      u_t^a = u_t^b + \mathbf{K}\bigl(y_t - H(u_t^b)\bigr),
    \end{equation}
    where $H$ is the observation operator (e.g., spatial averaging)
    and $\mathbf{K}$ is the Kalman gain that balances forecast and
    observation uncertainty.
\end{enumerate}

CASCADE-DD implements exactly this structure:
\begin{center}
\begin{tabular}{@{}lll@{}}
  \toprule
  \textbf{NWP concept} & \textbf{CASCADE-DD component} & \textbf{Equation} \\
  \midrule
  Previous analysis $u_{t-1}^a$
    & HR state $u_{t-1}$
    & --- \\
  Model forecast $u_t^b$
    & Advected state $u_{\mathrm{adv}} = \mathrm{warp}(u_{t-1}, \mathbf{v}_t)$
    & (\ref{eq:dd-adv}) \\
  Observation $y_t$
    & LR frame $\mathrm{LR}_t$
    & --- \\
  Observation operator $H$
    & $\mathrm{pool}(\cdot)$ (avg-pool to LR)
    & --- \\
  Innovation $y_t - H(u_t^b)$
    & $\mathrm{LR}_t - \mathrm{pool}(u_{\mathrm{adv}})$
    & (\ref{eq:dd-assim}) \\
  Kalman gain $\mathbf{K}$
    & AssimilationNet (learned, spatially varying)
    & (\ref{eq:dd-assim}) \\
  Analysis $u_t^a$
    & $u_{\mathrm{adv}} + \delta$
    & --- \\
  Subgrid parameterisation
    & ScaleRefine (SubgridNet loop)
    & (\ref{eq:dd-refine}) \\
  \bottomrule
\end{tabular}
\end{center}

The key difference from classical Kalman filtering is that the gain
$\mathbf{K}$ is not derived from an error covariance matrix but
\emph{learned end-to-end} from data.  This is both an advantage
(it can represent nonlinear, flow-dependent corrections that a linear
Kalman gain cannot) and a risk (it may overfit to the training
distribution).  The physics-based structure, innovation-driven
correction with bounded magnitude, constrains the learned gain to
behave sensibly even outside the training distribution.

\subsection{Temporal Dimension Addition}
\label{sec:temporal-physics}

The physics described above, advection, frontogenesis, scale
decomposition, data assimilation, all require \emph{temporal
information} to be useful.  A single snapshot cannot distinguish
a sharp front from a numerical artifact, or determine whether
a gradient should be intensified or smoothed.  The temporal
sequence resolves these ambiguities:

\begin{enumerate}
  \item \textbf{Velocity estimation requires time.}
    The advection equation (\cref{eq:advection}) contains
    $\mathbf{v}$, the velocity field.  From a single frame,
    $\mathbf{v}$ is unobservable, there is no motion to measure.
    Two or more frames allow the FlowNet to estimate $\mathbf{v}$
    by tracking how features move between snapshots (learned optical
    flow).  This estimated velocity is the foundation for all
    downstream physics: the warp direction, the frontogenesis
    alignment, and the subgrid conditioning.

  \item \textbf{Frontogenesis is directional.}
    Sharpening a gradient isotropically (as a single-frame SR would)
    creates artifacts, ringing, halos, cross-gradient noise.
    The temporal flow $\mathbf{v}_t$ defines the deformation axis:
    the SubgridNet sharpens \emph{along the motion direction},
    which is the physically correct frontogenetic axis.  Without
    temporal context, the network has no way to determine this
    direction.

  \item \textbf{Sub-pixel displacement carries real information.}
    If a feature moves by 0.3 pixels between frames at the LR grid,
    consecutive observations sample the feature at different
    sub-pixel phases.  This is genuine information about the HR
    structure, the same principle used in multi-frame SR in
    satellite remote sensing.  A single frame contains no sub-pixel
    phase diversity.

  \item \textbf{The assimilation cycle requires a forecast to correct.}
    The innovation $y_t - H(u_t^b)$ is only meaningful when there
    is a forecast $u_t^b$ to compare against.  That forecast comes
    from advecting the previous HR state forward in time, which
    requires knowing the temporal sequence.  Without time-stepping,
    there is no forecast, no innovation, and no assimilation; each
    frame would be super-resolved independently, losing the
    accumulated fine-scale information.

  \item \textbf{Temporal coherence distinguishes physics from artifacts.}
    A sharp feature that persists and moves consistently across frames
    is physical.  A sharp feature that appears in one frame and
    vanishes in the next is noise.  By evolving the HR state through
    time via transport, CASCADE-DD enforces that fine-scale detail
    must be temporally consistent with the observed motion, a
    powerful regularisation that no single-frame method can provide.
\end{enumerate}

\paragraph{Summary: each component needs time.}
\begin{center}
\small
\begin{tabular}{@{}lll@{}}
  \toprule
  \textbf{Physical mechanism} & \textbf{Network component}
    & \textbf{What time provides} \\
  \midrule
  Advection (\cref{eq:advection})
    & FlowNet
    & Velocity $\mathbf{v}_t$ from inter-frame displacement \\
  Frontogenesis (\cref{eq:frontogenesis})
    & SubgridNet
    & Deformation axis (direction to sharpen) \\
  Sub-pixel sampling
    & FlowNet
    & Phase diversity from feature displacement \\
  Data assimilation (\cref{eq:kalman})
    & AssimilationNet
    & Innovation from forecast--observation mismatch \\
  Conservation (\cref{eq:continuity})
    & Warp + LR loss
    & Accumulated HR state that persists across frames \\
  \bottomrule
\end{tabular}
\end{center}

Without temporal sequences, CASCADE reduces to a single-frame
scale-refinement loop better than a plain CNN thanks to the
transport inductive bias, but unable to leverage the advective,
frontogenetic, and assimilative physics that give the framework its
full power.

\section{Results and Discussion}
\label{sec:results}

We evaluate CASCADE on the SEVIR dataset using the Vertically Integrated
Liquid (VIL) radar product, a proxy for convective intensity and storm
structure.  SEVIR provides spatiotemporal sequences of severe weather
events, making it a natural testbed for methods that aim to
super-resolve evolving precipitation systems.  We report both continuous
image metrics (MSE, MAE, RMSE, PSNR, SSIM) and threshold-based skill
scores for light, moderate, and heavy regimes (CSI, HSS, POD, FAR, and
bias).

Continuous image metrics assess pixel-wise fidelity over the full field.
MSE and RMSE penalise large errors (and are therefore sensitive to missed
high-VIL cores), while MAE is more robust and reflects typical absolute
deviations. PSNR is a logarithmic re-scaling of MSE commonly used in SR;
higher PSNR indicates lower overall reconstruction error. SSIM measures
structural similarity (local contrast and spatial patterns) and is
important because convective organisation is not captured by amplitude
error alone.

\begin{figure}[t]
  \centering
  \includegraphics[width=0.49\linewidth]{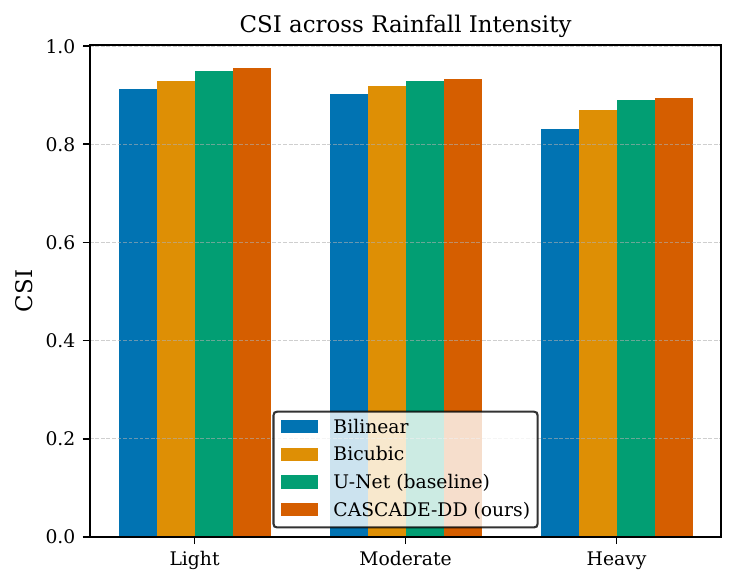}\hfill
  \includegraphics[width=0.49\linewidth]{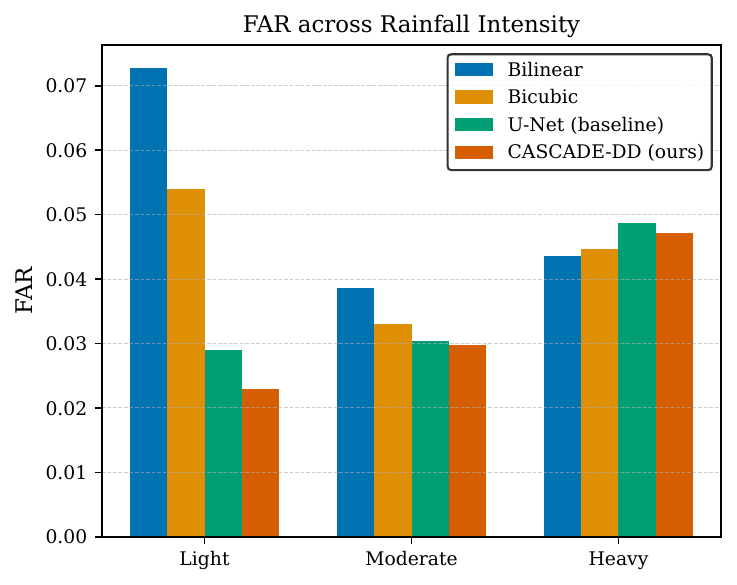}\\[0.5em]
  \includegraphics[width=0.49\linewidth]{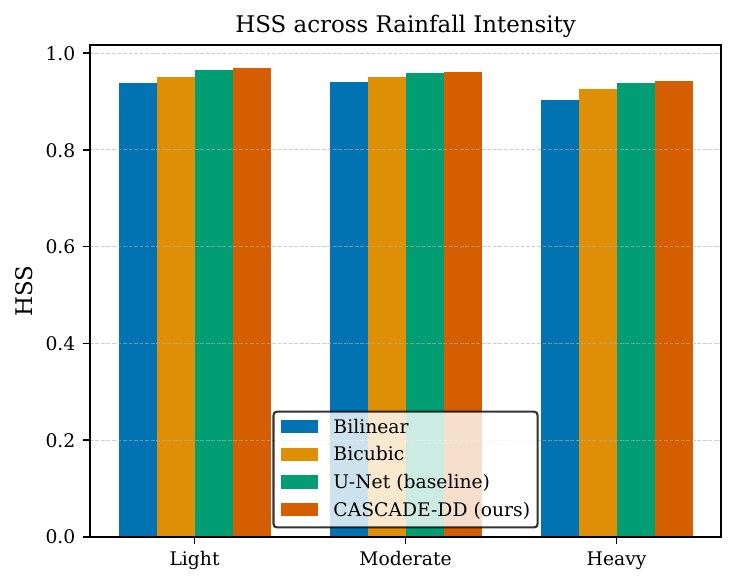}\hfill
  \includegraphics[width=0.49\linewidth]{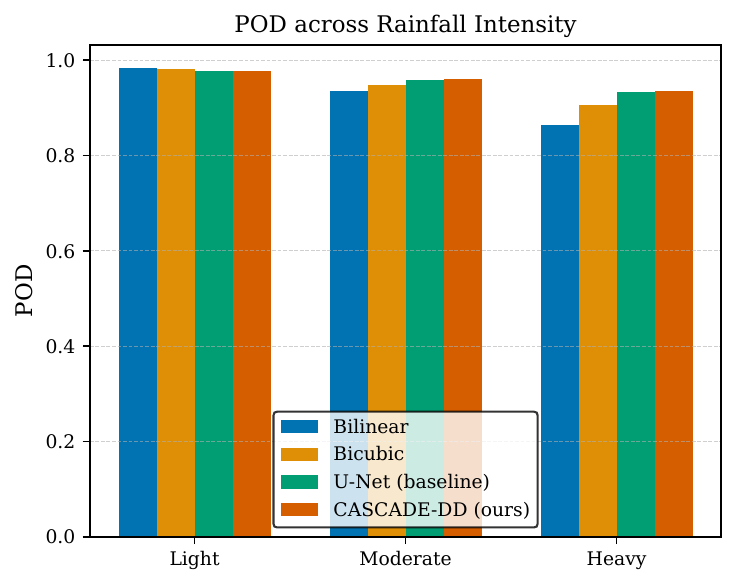}
  \caption{Varification metrics: CSI (top left), FAR (top right), HSS (bottom left), POD (bottom right).}
  \label{fig:skill-metrics-2x2}
\end{figure}

Threshold-based verification complements these averages by focusing on
event detection at scientifically relevant intensities. For a given VIL
threshold defining a regime (light, moderate, heavy), the critical
success index (CSI) measures overlap of predicted and observed exceedance
regions (balancing misses and false alarms). The Heidke skill score (HSS)
measures categorical skill relative to chance and is less sensitive to
the large number of correct negatives in sparse-event settings.
Probability of detection (POD) quantifies how many observed exceedances
are recovered, false alarm ratio (FAR) quantifies spurious exceedances,
and bias measures whether the model systematically over- or under-predicts
the frequency of threshold exceedance. Reporting both classes of metrics
is essential in convective SR: a model can achieve low MAE/PSNR by
smoothing extremes, yet still fail to detect intense cores and coherent
storm structures.

Table~\ref{tab:sevir-metrics} summarizes the main results. CASCADE-DD
outperforms a strong U-Net baseline across pixel-level and event-based
scores, despite the U-Net being substantially larger. In addition to
accuracy, CASCADE provides interpretable diagnostics through its learned
resolved flow ($\mathbf{v}_t$), subgrid transport ($\mathbf{v}_s$), and
assimilation correction ($\delta$). These diagnostics enable qualitative
inspection of how fine-scale structure is transported and where
observation-driven updates occur.

\begin{table}[t]
  \centering
  \caption{Quantitative evaluation of 4$\times$ super-resolution on SEVIR VIL radar data. Best results in \textbf{bold}.}
  \label{tab:sevir-metrics}
  \small
  \setlength{\tabcolsep}{3pt}
  \begin{tabular}{lccccccc}
    \toprule
    Method & PSNR$\uparrow$ & SSIM$\uparrow$ & MAE$\downarrow$ & CSI$_\mathrm{L}\uparrow$ & CSI$_\mathrm{M}\uparrow$ & CSI$_\mathrm{H}\uparrow$ & HSS$_\mathrm{M}\uparrow$ \\
    \midrule
    Bilinear & 31.70 & 0.9336 & 0.01019 & 0.913 & 0.902 & 0.831 & 0.941 \\
    Bicubic & 33.56 & 0.9509 & 0.00814 & 0.929 & 0.919 & 0.869 & 0.951 \\
    U-Net (baseline) & 35.33 & 0.9637 & 0.00676 & 0.949 & 0.930 & 0.891 & 0.958 \\
    CASCADE-DD (ours) & \textbf{35.88} & \textbf{0.9693} & \textbf{0.00616} & \textbf{0.955} & \textbf{0.933} & \textbf{0.895} & \textbf{0.960} \\
    \bottomrule
  \end{tabular}
\end{table}

\subsection{Advantages over Classical SR Methods}
\label{sec:advantages}

Classical super-resolution methods, bicubic interpolation, sparse-coding
SR, SRCNN, EDSR, and GAN-based approaches (SRGAN, ESRGAN), treat each
frame as an independent image and reconstruct high-frequency detail via
learned pixel mappings.  The CASCADE framework differs in several
structurally important ways.

\subsubsection{Transport-based reconstruction instead of pixel hallucination}

Classical deep SR networks learn a direct mapping
$f_\theta: \mathbb{R}^{H\times W} \to \mathbb{R}^{rH\times rW}$
whose decoder must synthesise fine-grained texture from the latent space.
In geophysical data, much of the ``missing'' detail at fine scales is not
random texture but \emph{structure that has been spatially averaged away}:
sharp fronts, convective cells, orographic precipitation bands.

CASCADE recovers this structure by \emph{transporting} intensity along
learned velocity fields (\cref{eq:scale-step}).  The SubgridNet does not
hallucinate new pixel values; it predicts \emph{where existing values
should move to} at the finer grid.  This is analogous to the physical
process of frontogenesis, where large-scale deformation concentrates
gradients into sharp fronts, the information is already present in the
coarse field, it just needs to be ``compressed'' spatially.

\begin{figure}[t]
  \centering
  \includegraphics[width=\linewidth]{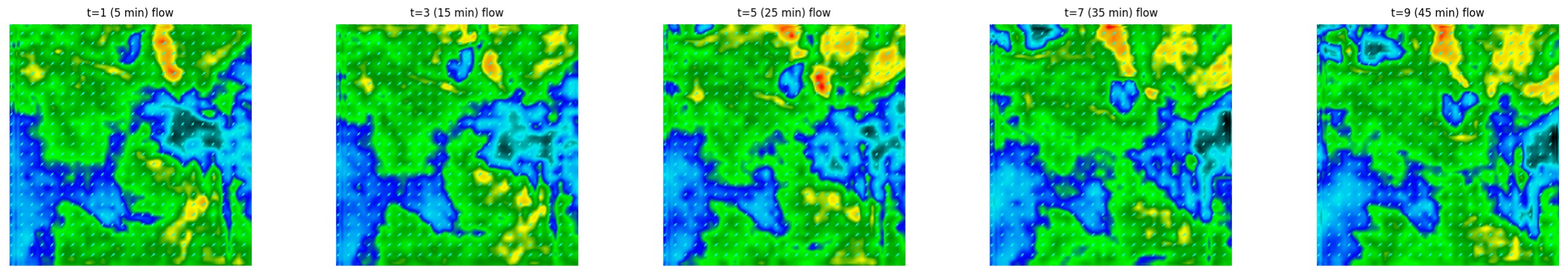}
  \caption{Leared storm advection flow.}
  \label{fig:storm-advection-flow}
\end{figure}

\paragraph{Consequences:}
\begin{itemize}
  \item Sharper outputs without adversarial training or perceptual losses
        that risk introducing physically implausible artifacts.
  \item Mass (or energy, moisture) is approximately conserved because
        warping redistributes rather than creates intensity, and this is
        further enforced by the LR-consistency constraint
        $\|\mathrm{pool}(\hat{u}_{\mathrm{HR}}) - u_{\mathrm{LR}}\|^2$.
  \item The velocity-smoothness regularisation
        $\|\nabla \mathbf{v}\|^2$ prevents unphysically noisy velocity
        fields, acting as a learned analogue of a viscosity constraint.
\end{itemize}

\subsubsection{Explicit separation of resolved and subgrid dynamics}

The architecture enforces a \emph{scale decomposition}: the FlowNet
captures large-scale, resolved motion (wind advection, synoptic
propagation), while the SubgridNet captures the residual fine-scale
transport that cannot be resolved at the LR grid.  This mirrors the
Reynolds decomposition $\mathbf{u} = \bar{\mathbf{u}} + \mathbf{u}'$
used in turbulence modelling and subgrid parameterisation in climate
models.

Classical SR has no such decomposition, the single decoder must
simultaneously handle both the large-scale structure and fine-scale
texture, making it harder to generalise when the balance between
resolved and unresolved scales changes (e.g., different seasons, climate
regimes, or forecast lead times).

\subsubsection{Built-in physical constraints}

\begin{itemize}
  \item \textbf{LR-consistency loss:}
    $\mathcal{L}_{\mathrm{LR}} = \|\mathrm{pool}(\hat{u}_{\mathrm{HR}})
    - u_{\mathrm{LR}}\|^2$ guarantees that the SR output, when coarsened
    back to the input resolution, reproduces the original observation.
    Classical SR does not enforce this, so the network can drift from the
    input signal.
  \item \textbf{Velocity smoothness:}
    $\mathcal{L}_{\mathrm{smooth}} = \|\nabla \mathbf{v}\|^2$ acts as a
    soft Navier--Stokes-like constraint on the transport field, preventing
    discontinuous or noisy displacements.
  \item \textbf{Conservation by construction:}
    Semi-Lagrangian warp is a \emph{redistribution} operation, not a
    generative one.  The total integral of the field changes only due to
    boundary effects (padding), not due to the network inventing new
    intensity.
\end{itemize}

\subsubsection{Interpretability}

The learned velocity fields $\mathbf{v}_t$ and $\mathbf{v}_s$ can be
visualised and physically interpreted:
\begin{itemize}
  \item $\mathbf{v}_t$ (resolved flow): should align with the true
        large-scale wind or propagation velocity, this can be verified
        against reanalysis wind fields.
  \item $\mathbf{v}_s$ (subgrid flow): should point toward gradient
        maxima, tightening fronts and intensifying local extremes, the
        scale-transport signature of frontogenesis.
  \item $\delta$ (assimilation correction in CASCADE-DD): shows spatially
        where the advected forecast disagrees with observations, analogous
        to the innovation field in Kalman filtering.
\end{itemize}
No equivalent diagnostic exists for a standard SRCNN/ESRGAN decoder.

\section{Conclusion}
\label{sec:conclusion}

We have introduced CASCADE, a physics-informed framework that reframes 
geophysical super-resolution as an explicit transport process across scales. 
Rather than hallucinating fine-scale structure per pixel, CASCADE recovers 
detail by iteratively advecting coarse information along learned velocity 
fields through semi-Lagrangian warping. By hard-wiring advection as the core 
operator, decomposing dynamics into resolved and subgrid components, and 
incorporating assimilation-style innovation updates, CASCADE encodes fundamental 
physical principles directly into the architecture. Evaluation on SEVIR radar 
data demonstrates that CASCADE-DD outperforms strong U-Net baselines across 
both continuous metrics (PSNR, SSIM, MAE) and threshold-based verification 
scores (CSI, HSS, POD) while providing interpretable diagnostics through 
visualizable velocity and correction fields unavailable from black-box decoders.

The transport-centric design addresses key limitations of classical SR methods 
when applied to physical data: CASCADE produces temporally coherent outputs by 
construction (via state advection), approximately conserves mass and energy 
(warping redistributes rather than creates intensity), and enforces consistency 
with coarse observations through explicit LR-pooling loss. The learned scale 
decomposition mirrors the closure problem in numerical weather prediction, 
allowing the model to automatically separate large-scale resolved motion from 
unresolved subgrid transport. This structural approach reduces what must be 
learned from data and improves physical plausibility compared to purely 
data-driven regression.

Several limitations suggest directions for future work. CASCADE assumes 
fine-scale structure can be recovered by transport, which holds well for 
advection-dominated fields but may be less appropriate for phenomena arising 
spontaneously without large-scale precursors (e.g., random convective initiation). 
Extension to multivariate downscaling requires maintaining physical relationships 
between coupled fields. As a deterministic model, CASCADE lacks uncertainty 
quantification; ensemble-based or generative extensions could provide probability 
distributions while preserving transport consistency. Finally, like all 
data-driven methods, generalization to new regions, climate regimes, or extreme 
events outside the training distribution remains an open challenge requiring 
domain adaptation techniques or stronger physics constraints.

Beyond atmospheric super-resolution, CASCADE exemplifies a broader principle 
in scientific machine learning: when governing physics is known, encode it 
architecturally rather than hoping networks rediscover it from data. This 
"physics in the architecture" paradigm: hard-wiring transport, conservation, 
and temporal evolution as structural components—reduces sample complexity, 
improves interpretability, and enhances robustness compared to black-box 
approaches. As ML-based weather and climate models approach operational 
deployment, CASCADE demonstrates that physically consistent, temporally coherent, 
and interpretable downscaling is achievable when the right operators are made 
explicit in the model design rather than hidden in learned weights.

\bibliography{references}

@article{lam2023graphcast,
  title        = {GraphCast: Learning skillful medium-range global weather forecasting},
  author       = {Lam, Remi and Sanchez-Gonzalez, Alvaro and others},
  journal      = {Science},
  year         = {2023},
}

@article{bi2023panguweather,
  title        = {Pangu-Weather: A {3D} high-resolution model for fast and accurate global weather forecasting},
  author       = {Bi, Kaifeng and others},
  journal      = {Nature},
  year         = {2023},
}

@article{pathak2022fourcastnet,
  title        = {FourCastNet: A global data-driven high-resolution weather model using adaptive Fourier neural operators},
  author       = {Pathak, Jaideep and others},
  journal      = {Proceedings of the National Academy of Sciences},
  year         = {2022},
}

@inproceedings{nguyen2023climax,
  title        = {ClimaX: A foundation model for weather and climate},
  author       = {Nguyen, Tung and others},
  booktitle    = {International Conference on Machine Learning (ICML)},
  year         = {2023},
}

@article{merchant2023metnet3,
  title        = {MetNet-3: A state-of-the-art neural weather model for precipitation nowcasting},
  author       = {Merchant, Sam and others},
  journal      = {arXiv},
  year         = {2023},
  note         = {arXiv preprint},
}

@article{karniadakis2021pinns,
  title        = {Physics-informed machine learning},
  author       = {Karniadakis, George Em and others},
  journal      = {Nature Reviews Physics},
  year         = {2021},
}

@article{cuomo2022scientificml,
  title        = {Scientific machine learning through physics-informed neural networks: Where we are and what{'}s next},
  author       = {Cuomo, Salvatore and others},
  journal      = {Journal of Scientific Computing},
  year         = {2022},
}

@inproceedings{lim2017edsr,
  title        = {Enhanced Deep Residual Networks for Single Image Super-Resolution},
  author       = {Lim, Bee and Son, Sanghyun and Kim, Heewon and Nah, Seungjun and Mu Lee, Kyoung},
  booktitle    = {IEEE Conference on Computer Vision and Pattern Recognition Workshops (CVPRW)},
  year         = {2017},
}

@inproceedings{zhang2018rcan,
  title        = {Image Super-Resolution Using Very Deep Residual Channel Attention Networks},
  author       = {Zhang, Yulun and Li, Kunpeng and Li, Kai and Wang, Lichen and Zhong, Bineng and Fu, Yun},
  booktitle    = {European Conference on Computer Vision (ECCV)},
  year         = {2018},
}

@inproceedings{wang2018esrgan,
  title        = {{ESRGAN}: Enhanced Super-Resolution Generative Adversarial Networks},
  author       = {Wang, Xintao and Yu, Ke and Wu, Shixiang and Gu, Jinjin and Liu, Yihao and Dong, Chao and Qiao, Yu and Loy, Chen Change},
  booktitle    = {European Conference on Computer Vision Workshops (ECCVW)},
  year         = {2018},
  note         = {arXiv preprint arXiv:1809.00219},
}

@article{vandal2017deepsd,
  title        = {{DeepSD}: Generating High Resolution Climate Change Projections through Single Image Super-Resolution},
  author       = {Vandal, Thomas and Kodra, Evan and Ganguly, Saptarshi and Michaelis, Andrew and Nemani, Ramakrishna and Ganguly, Auroop R.},
  journal      = {Proceedings of the 23rd ACM SIGKDD International Conference on Knowledge Discovery and Data Mining (KDD)},
  year         = {2017},
}

@article{huang2023systematicreview,
  title        = {A systematic review of deep learning for statistical downscaling of climate variables},
  author       = {Huang, {X.} and others},
  journal      = {Environmental Research Letters},
  year         = {2023},
}

@article{wan2024genbcsr,
  title        = {Generative Bias-Corrected Super-Resolution for Climate and Weather Fields},
  author       = {Wan, {Z.} and others},
  journal      = {arXiv preprint},
  year         = {2024},
}

@article{li2020fno,
  title        = {Fourier Neural Operator for Parametric Partial Differential Equations},
  author       = {Li, Zongyi and Kovachki, Nikola and Azizzadenesheli, Kamyar and Liu, Burigede and Bhattacharya, Kaushik and Stuart, Andrew and Anandkumar, Anima},
  journal      = {arXiv preprint arXiv:2010.08895},
  year         = {2020},
}

@article{li2021pino,
  title        = {Physics-Informed Neural Operator for Learning Partial Differential Equations},
  author       = {Li, Zongyi and others},
  journal      = {arXiv preprint arXiv:2111.03794},
  year         = {2021},
}

@article{azizzadenesheli2024neuraloperators,
  title        = {Neural Operators: A Review},
  author       = {Azizzadenesheli, Kamyar and others},
  journal      = {arXiv preprint},
  year         = {2024},
}

@article{leinonen2023ldmnowcasting,
  title        = {Latent diffusion models for generative precipitation nowcasting with accurate uncertainty quantification},
  author       = {Leinonen, Jussi and Hamann, Ulrich and Nerini, Daniele and Germann, Urs and Franch, Gabriele},
  journal      = {arXiv preprint arXiv:2304.12891},
  year         = {2023},
}

@article{ling2024srndiff,
  title        = {{SRNDiff}: Short-term Rainfall Nowcasting with Condition Diffusion Model},
  author       = {Ling, Xudong and Li, Chaorong and Qin, Fengqing and Yang, Peng and Huang, Yuanyuan},
  journal      = {arXiv preprint arXiv:2402.13737},
  year         = {2024},
}

@article{wen2024duocast,
  title        = {{DuoCast}: Duo-Probabilistic Meteorology-Aware Model for Extended Precipitation Nowcasting},
  author       = {Wen, Penghui and Bai, Lei and He, Mengwei and Filippi, Patrick and Zhang, Feng and Bishop, Thomas Francis and Wang, Zhiyong and Hu, Kun},
  journal      = {arXiv preprint arXiv:2412.01091},
  year         = {2024},
}

@article{foo2025stldm,
  title        = {{STLDM}: Spatio-Temporal Latent Diffusion Model for Precipitation Nowcasting},
  author       = {Foo, Shi Quan and Wong, Chi-Ho and Gao, Zhihan and Yeung, Dit-Yan and Wong, Ka-Hing and Wong, Wai-Kin},
  journal      = {arXiv preprint arXiv:2512.21118},
  year         = {2025},
}

\section*{Supplementary Information}

\subsection*{When to use CASCADE?}
\label{sec:when}

The CASCADE framework is most useful when the following conditions hold:

\begin{enumerate}[label=(\roman*)]
  \item \textbf{The fine-scale structure is governed by transport.}
    The approach assumes that high-resolution detail can be recovered
    by moving (warping) existing coarse-scale information, rather than
    by generating entirely new patterns.  This is a strong assumption
    that holds well for advection-dominated fields (wind, temperature
    fronts, moisture transport) but may not hold for fields where
    fine-scale structure is created \emph{in situ} without a
    coarse-scale precursor (e.g., random convective initiation over
    flat terrain).

  \item \textbf{Paired LR--HR training data exists.}
    The model requires co-registered LR and HR fields for supervised
    training.  In climate applications this typically means:
    \begin{itemize}
      \item HR: convection-permitting simulations or high-resolution
            analyses (e.g., HRES, CERRA, HCLIM at 2--5\,km).
      \item LR: coarsened versions of the above, or directly from
            reanalysis products (ERA5 at 0.25$^\circ$, $\sim$31\,km).
    \end{itemize}

  \item \textbf{The fields are approximately continuous and
    differentiable.}
    Semi-Lagrangian advection (bilinear interpolation in a grid-sample
    operation) assumes smooth fields.  Discrete or categorical data
    (land-use type, cloud mask) would require modifications.

  \item \textbf{Temporal sequences are available (for the full benefit).}
    While CASCADE-SR can in principle operate on a single frame
    ($N=1$, with the FlowNet receiving only one channel), the
    temporal advantage disappears.  The method is most powerful when
    consecutive observations at cadences shorter than the advective
    time scale are available, typically hourly to 6-hourly for
    synoptic-scale features.

  \item \textbf{Physical interpretability is valued.}
    If the application requires not just a ``better-looking'' output
    but also diagnostics about what the model is doing (flow fields,
    assimilation increments, conservation diagnostics), the CASCADE
    framework provides these natively.
\end{enumerate}

\subsection*{Specific Application Cases}
\label{sec:cases}

\subsubsection*{Case 1: ERA5 to convection-permitting resolution
            (statistical--dynamical downscaling)}

\begin{tabular}{@{}ll@{}}
  \toprule
  \textbf{LR input}  & ERA5 reanalysis, 0.25$^\circ$ ($\sim$31\,km),
                        6-hourly or hourly \\
  \textbf{HR target}  & CERRA (5.5\,km) or COSMO/HCLIM runs (2--3\,km) \\
  \textbf{Variables}   & 2\,m temperature, 10\,m wind, precipitation,
                        MSLP, geopotential \\
  \textbf{Architecture} & CASCADE-DD (Problem~B) \\
  \bottomrule
\end{tabular}

\medskip
The time-stepping loop evolves an HR state driven by ERA5 at every
analysis time.  FlowNet learns the synoptic advection between 6-hourly
or hourly snapshots; AssimilationNet corrects the forecast toward the
coarse observation; SubgridNet adds orographic precipitation enhancement,
sea-breeze fronts, and convective-scale structure that ERA5 cannot
resolve.

\paragraph{Advantage over classical SR:}
Classical bicubic or EDSR-style downscaling of ERA5 would treat each
time step independently.  A cold front moving across complex terrain
would be sharpened isotropically at each snapshot, with no guarantee of
temporal coherence.  CASCADE-DD advects the sharpened front forward and
only corrects where the LR observation disagrees, producing a
physically consistent HR trajectory.

\subsubsection*{Case 2: Satellite imagery enhancement
            (geostationary temporal SR)}

\begin{tabular}{@{}ll@{}}
  \toprule
  \textbf{LR input}  & Geostationary IR/WV channels (4--8\,km pixel),
                        15-minute cadence \\
  \textbf{HR target}  & LEO satellite passes (0.5--1\,km) or
                        radar composites \\
  \textbf{Variables}   & Brightness temperature, cloud-top height,
                        precipitation rate \\
  \textbf{Architecture} & CASCADE-SR (Problem~A) with $N=4$--$8$ frames \\
  \bottomrule
\end{tabular}

\medskip
High temporal cadence of geostationary satellites provides dense motion
information.  FlowNet extracts cloud motion vectors (analogous to
AMVs); SubgridNet sharpens cloud edges along the motion direction.
This is directly useful for nowcasting and convective initiation
detection, where the exact position and sharpness of a developing cloud
cluster matters.

\subsubsection*{Case 3: Ocean model downscaling
            (mesoscale eddy reconstruction)}

\begin{tabular}{@{}ll@{}}
  \toprule
  \textbf{LR input}  & Global ocean model (NEMO, MOM6) at 1/4$^\circ$,
                        daily \\
  \textbf{HR target}  & Eddy-resolving simulations at 1/12$^\circ$ or
                        satellite altimetry \\
  \textbf{Variables}   & SSH, SST, surface currents \\
  \textbf{Architecture} & CASCADE-DD (Problem~B) \\
  \bottomrule
\end{tabular}

\medskip
Mesoscale eddies are transport-dominated structures: they propagate
westward, interact, merge, and filament.  A coarse ocean model partially
resolves the largest eddies but smears the sharp SST fronts at their
boundaries.  CASCADE-DD can maintain a high-resolution SSH/SST state,
advect it with the learned mesoscale flow, assimilate against the daily
coarse model output, and refine submesoscale filaments via the SubgridNet.

\paragraph{Key advantage:}
Conservation of tracer mass is critical in ocean applications.  The warp
operation inherently conserves (up to boundary padding), and the
LR-consistency loss prevents the network from inventing water mass.

\subsubsection*{Case 4: Wind resource assessment
            (terrain-aware temporal downscaling)}

\begin{tabular}{@{}ll@{}}
  \toprule
  \textbf{LR input}  & ERA5 or GFS 10\,m / 100\,m wind, 0.25$^\circ$,
                        hourly \\
  \textbf{HR target}  & WRF or COSMO simulations at 1--3\,km, or
                        met-mast / lidar data \\
  \textbf{Variables}   & $u$, $v$ wind components at hub height \\
  \textbf{Architecture} & CASCADE-SR (Problem~A) or CASCADE-DD \\
  \bottomrule
\end{tabular}

\medskip
Wind fields over complex terrain develop channelling, lee waves, and
thermally driven flows (valley/slope winds) that are absent from ERA5.
FlowNet learns the synoptic-scale wind evolution; SubgridNet adds the
terrain-induced perturbations.  Temporal sequences capture the diurnal
cycle of thermally driven flows---a single snapshot cannot distinguish
morning upslope from afternoon downslope flow, but a 6--12 hour window
resolves the cycle.

\subsubsection*{Case 5: Precipitation downscaling for hydrology}

\begin{tabular}{@{}ll@{}}
  \toprule
  \textbf{LR input}  & ERA5 or GPM-IMERG precipitation, 0.1--0.25$^\circ$,
                        hourly \\
  \textbf{HR target}  & Radar-based QPE (1--4\,km) or gauge-adjusted
                        analyses \\
  \textbf{Variables}   & Precipitation rate, accumulated precipitation \\
  \textbf{Architecture} & CASCADE-DD (Problem~B) \\
  \bottomrule
\end{tabular}

\medskip
Precipitation is the most challenging variable because it is
spatially intermittent and highly non-Gaussian.  However, precipitating
systems \emph{move}---squall lines propagate, mesoscale convective
systems have well-defined motion---and the temporal structure is rich.
FlowNet can track storm motion across consecutive LR frames;
SubgridNet can sharpen the precipitation core and trailing stratiform
region; AssimilationNet corrects when new convective cells initiate
that were not present in the advected forecast.

\paragraph{Caveat:}
For the most intermittent, smallest-scale convective precipitation, the
transport assumption weakens.  A hybrid approach combining CASCADE with a
conditional generative component for stochastic initiation may be
needed.

\subsubsection*{Case 6: Climate projection downscaling
            (GCM to regional scale)}

\begin{tabular}{@{}ll@{}}
  \toprule
  \textbf{LR input}  & CMIP6 GCM output (1--2$^\circ$), daily or
                        6-hourly \\
  \textbf{HR target}  & CORDEX regional simulations (12--25\,km) or
                        observations \\
  \textbf{Variables}   & Temperature, wind, precipitation, humidity \\
  \textbf{Architecture} & CASCADE-DD (Problem~B) \\
  \bottomrule
\end{tabular}

\medskip
Long climate projections at coarse resolution need downscaling to assess
regional impacts (flood risk, crop yield, energy demand).  Classical
statistical downscaling (quantile mapping, analogue methods) is
frame-by-frame and cannot capture the temporal sequencing of events
(e.g., a multi-day heat wave's diurnal structure).  CASCADE-DD
time-steps through the GCM output, maintaining a physically coherent
HR trajectory that respects both the GCM's large-scale evolution and
the learned fine-scale physics.

\paragraph{Open question:}
Generalization to unseen climate states (future warming) requires that
the learned subgrid closure extrapolates beyond the training
distribution.  Physics-based regularisation (conservation, smoothness,
flow-gradient coupling) may help, but this remains an active research
challenge for all ML-based downscaling approaches.

\subsection*{Summary}

\begin{table}[h]
\centering
\small
\begin{tabular}{@{}p{4cm}p{5.5cm}p{5.5cm}@{}}
  \toprule
  & \textbf{Classical deep SR}
  & \textbf{CASCADE} \\
  \midrule
  Core operation
    & Per-pixel regression
    & Transport (semi-Lagrangian warp) \\
  Temporal information
    & None (single frame)
    & $N$-frame flow estimation / time-stepping \\
  Conservation
    & Not enforced
    & Approximate by construction + LR loss \\
  Interpretability
    & Black-box decoder
    & Visualisable velocity fields + innovation \\
  Temporal coherence
    & Not guaranteed (flickering)
    & Built-in via state advection \\
  Architecture bias
    & Translation equivariance (CNN)
    & Transport equivariance + scale decomposition \\
  \bottomrule
\end{tabular}
\caption{Structural comparison between classical deep SR and the CASCADE
  framework.}
\end{table}

The CASCADE framework reframes super-resolution as a \emph{physical
transport process across scales}, rather than a pixel-level regression
task.  By separating resolved flow from subgrid closure and, in the
dynamical downscaling variant, incorporating data assimilation within
a time-stepping loop, the approach provides temporally coherent,
physically constrained, and interpretable high-resolution
reconstructions.  Its natural domain of application is advection-dominated
geophysical fields where paired LR--HR data and temporal sequences are
available.

This transport-centric design is particularly valuable for extreme events:
it targets the coherent evolution of localized, high-intensity features that
dominate impacts, while reducing the risk of frame-to-frame flicker or
unphysical creation/loss of mass that can misrepresent hazard magnitude.

This transport-centric design is particularly valuable for extreme events:
it targets the coherent evolution of localized, high-intensity features that
dominate impacts, while reducing the risk of frame-to-frame flicker or
unphysical creation/loss of mass that can misrepresent hazard magnitude.

\end{document}